\documentclass[aps,prl,twocolumn,showpacs]{revtex4}

\newcommand{\swap}{$\sqrt{\mbox{{\footnotesize SWAP}}}$ \ }
\newcommand{\Jeff}{J_{\mbox{\footnotesize eff}}}
\newcommand{\Jheis}{J_{\mbox{\footnotesize H}}}

\usepackage{graphicx}

\begin{document}

\title{Quantum control and entanglement
using periodic driving fields}

\author{C.E.~Creffield}
\affiliation{Department of Physics and Astronomy, University College London,
Gower Street, London WC1E 6BT, United Kingdom}

\date{\today}

\pacs{03.67.Mn, 03.75.Lm}

\begin{abstract}
We propose a scheme for producing directed
motion in a lattice system by applying a periodic driving potential. 
By controlling the dynamics by means of
the effect known as coherent destruction of tunneling, we demonstrate
a novel ratchet-like effect that enables particles to be
coherently manipulated and steered without requiring local control.
Entanglement between particles can also be controllably generated,
which points to the attractive possibility
of using this technique for quantum information processing.
\end{abstract}

\maketitle

{\em Introduction}
Controlling the time-evolution of quantum states, and engineering
entanglement between quantum particles, are two of the major
tasks required for quantum information processing. Although
the majority of experimental demonstrations of entanglement
distribution have so far employed photons, future practical 
implementations of quantum computers will almost certainly be 
based on condensed matter systems. Efforts in this direction include
controlling electronic charge or spin (``spintronic'') degrees
of freedom in coupled quantum dots \cite{qdots},
manipulating the dynamics of Josephson junctions \cite{jjunction},
and using spin chains \cite{chains} as quantum
communication channels. Recently bosons held in optical lattice
potentials have also been suggested as possible candidates,
using controlled collisions \cite{jaksch_qc} or the dipole-dipole
interaction \cite{dipole} to manipulate the system. 

In this work we demonstrate how particles
in a lattice potential can be controlled 
by applying an oscillatory driving field. 
Control is achieved by using the effect
termed ``coherent destruction of tunneling'' (CDT) \cite{hanggi}
to generate a ratchet-like motion. Unlike the majority of ratchets
which depend on the presence of dissipation to operate, this motion
results solely from the breaking of spatial and time symmetries in
the system, and so preserves the quantum coherence of the system.
Using this scheme, a pair of particles can be brought together
and allowed to interact -- thereby becoming 
entangled -- and then subsequently separated again to  
create entanglement between distinct spatial locations.
We are thus able to both selectively move and entangle
quantum particles using only the {\em global}
operation of varying the parameters of the driving field,
thus avoiding any need for the individual addressing of
lattice sites. 

\smallskip

{\em Model}
The specific physical system that we consider consists of ultracold
bosonic atoms, confined in a one-dimensional optical lattice potential 
created by the superposition of counter-propagating laser beams. 
This form of confinement provides an extremely clean and controllable 
lattice potential, and in addition their 
high degree of isolation from the environment gives these systems
rather long decoherence times, making them ideal for studying
quantum coherent phenomena.

The single-particle dynamics of the system can be described extremely 
well by the Hamiltonian \cite{jaksch}
\begin{equation}
H = \sum_{\langle i, j \rangle} \left[ J_{i j} \ 
a_i^{\dagger} a_j^{ } + \mbox{H.c.} \right] + 
K \sin \omega t \ \sum_{i} x_i n_i \ ,
\label{hamilton}
\end{equation}
where $a_i^{ } / a_i^{\dagger}$ are the standard bosonic 
destruction/creation operators and the tunneling
amplitudes $J_{i ,j}$ connect nearest-neighbor
sites $( i,j )$. Without loss of generality we shall
henceforth take $J_{i j}=J$. The amplitude and frequency
of the time-dependent driving field are described
by the parameters $K$ and $\omega$,
and $x_i$ is the spatial location of the $i$th
lattice site. This form of linear potential
has already been used in cold atom experiments
\cite{raizen} to induce CDT, and
can be straightforwardly implemented in an optical lattice
by introducing a periodic phase-modulation to one of the 
laser fields providing the standing wave potential.

Since the Hamiltonian (\ref{hamilton}) is
invariant under discrete translations in time of the drive-period,
$H(t) = H(t + n T)$, the Floquet theorem
allows us to write solutions of the Schr\"odinger equation
as $|\phi_n(t) \rangle = |u_n(t) \rangle \exp
\left[-i t \epsilon_n \right]$, where $| u_n(t) \rangle$ is
a $T$-periodic function called the Floquet function, and $\epsilon_n$
is termed the quasienergy. 
When two quasienergies approach degeneracy,
the timescale to observe tunneling between
the associated Floquet states diverges, and accordingly
the tunneling between them appears suppressed. In the limit
of high-frequency (when $\omega$ is the dominant energy-scale 
of the problem) it may be indeed shown 
\cite{hanggi,dunlap,holthaus,creff_pert} that the driven system
behaves like the undriven one, but with renormalized tunneling  
amplitudes. For sinusoidal driving this renormalization
takes the form 
$\Jeff = J {\cal J}_0 (K x/ \omega)$, where ${\cal J}_0$
is the zeroth Bessel function of the first kind
and $x$ is the intersite separation.
Thus when $K x/\omega$ is equal to a zero of ${\cal J}_0$
the system's tunneling dynamics are frozen, producing CDT.

\begin{center}
\begin{figure}
\includegraphics[width=0.40\textwidth,clip=true]{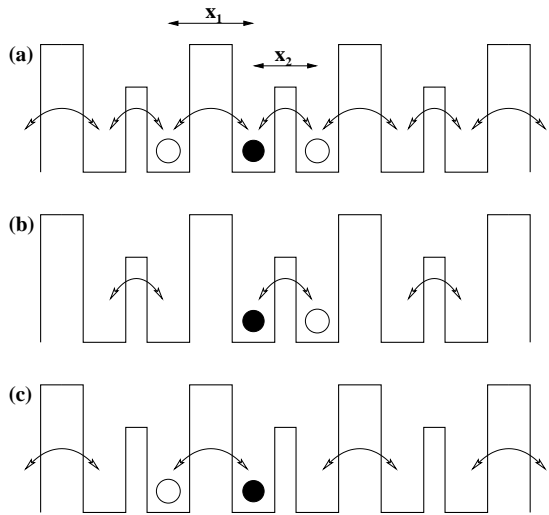}
\caption{We consider a bipartite lattice with two spacings: $x_1$ and
$x_2$. Permitted tunneling processes are shown by the arrows.
(a) In the absence of a driving field, a particle initialized in
a given lattice site (the filled circle) can tunnel to either
of its neighbors (empty circles). From there it can spread
over the entire lattice.
(b) If the lattice is sinusoidally-driven with a field such that
${\cal J}_0(K x_1/\omega) = 0$ then the tunneling processes
between sites separated by $x_1$ are suppressed, and the particle
can only tunnel to its right neighbor.
(c) Conversely, if the driving field satisfies
${\cal J}_0(K x_2/\omega) = 0$, then tunneling is only permitted
between sites separated by $x_1$, and the particle can only
tunnel to its left neighbor.}
\label{array}
\end{figure}
\end{center}

This way of regulating the
tunneling between sites has been recently proposed
to control the Mott-insulator
transition in Bose-Einstein condensates \cite{eckardt,creffield}.
To obtain a ratchet
effect, however, it is necessary to distinguish between
motion to the left and motion to the right. This may be achieved
by noting that the argument of the Bessel function depends
on the potential difference between neighboring sites, and thus 
on their spatial separation. Accordingly we consider a bipartite
lattice of form $ABABAB$, as shown in Fig.\ref{array},  
in which the $AB$ separation is not equal to that between $BA$. 
One possible realization of this would be a chain
of coupled double-well potentials \cite{twodim}.
Consider placing a single particle in the center of this lattice
(Fig.\ref{array}a). In the absence of a driving field it will 
rapidly disperse by tunneling to both its neighbors. 
If, however, the system is driven by a high-frequency sinusoidal
potential such that ${\cal J}_0(K x_1/\omega) = 0$, then tunneling
between sites separated by $x_1$ is destroyed and the lattice
divides into a set of disconnected dimers $(AB)(AB)(AB)$.
In this case the particle is unable to spread over
the lattice, and is restricted to making a Rabi oscillation
(Fig.\ref{array}b) between its initial location and
its neighbor to the right. The frequency of this oscillation
is determined by the value of the renormalized tunneling
for this process $\Jeff = J {\cal J}_0 (K x_2/\omega)$,
which will in general be non-zero.
Conversely, if the parameters of the driving field are
chosen such that  ${\cal J}_0(K x_2/\omega) = 0$ then
the lattice dimerises as $(BA)(BA)(BA)$ as shown in
Fig.\ref{array}c, and the Rabi oscillation will occur between
the initial site and its neighbor to the left.

\smallskip

\begin{center}
\begin{figure}
\includegraphics[width=.4\textwidth,clip=true]{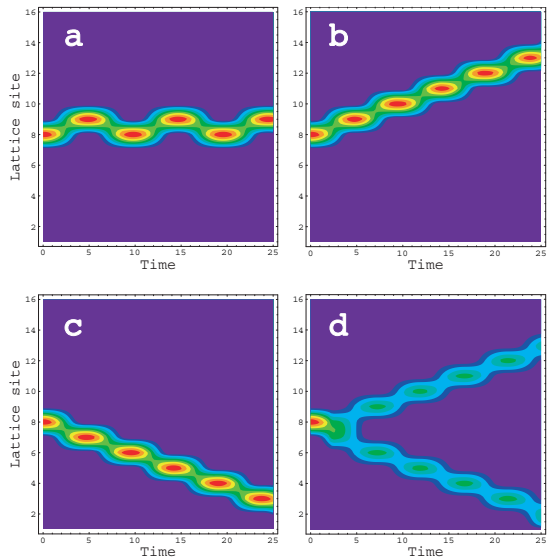}
\caption{Simulation of a single boson in a 16-site system.
(a) Under a periodic driving field of amplitude
$K x_1/\omega = 2.4048$ we obtain the situation
illustrated in Fig.\ref{array}b -- tunneling between sites
separated by $x_1$ is suppressed. Consequently the
particle makes a Rabi oscillation to just one of its neighbors.
(b) Under the driving field shown in Fig.\ref{field}
the $x_1$ and $x_2$ tunneling processes are periodically
opened and closed, producing a ratchet-like motion of the particle.
(c) Interchanging the order of the modulation of
the driving field produces motion in the opposite direction.
(d) Shortening the duration of the initial modulation
means that the first tunneling process will be incomplete.
If the initial modulation time is halved
the particle splits into two equal parts,
and under the driving field each part propagates in
different directions.}
\label{simulation}
\end{figure}
\end{center}

{\em Results}
To verify this effect, we show in Fig.\ref{simulation}a
the results of a numerical simulation of a
single particle in a 16-site system
with $x_1=1$ and $x_2=0.75$. The frequency of the driving is
set to a high value of $\omega=32 J$ to ensure that
the system is in the high-frequency regime, while its amplitude
satisfies $K x_1/ \omega=2.4048$ -- the first zero of ${\cal J}_0$.
As the analysis predicts, the particle indeed simply oscillates
between its initial location and one of its neighbors (that separated
by a distance of $x_2$), since tunneling between sites 
separated by $x_1$ has been suppressed.

The crispness of the Rabi oscillation immediately suggests a 
scheme to produce directed motion. If we
denote the period of this oscillation by $T_2$, then
at $t=T_2/2$ the particle has completely tunneled from its 
initial location $i$ to its neighbor $i+1$. If at this time the
parameters of the field are altered so that 
$K x_2/\omega=2.4048$, then this tunneling process
is suppressed and the particle instead begins to make
a Rabi oscillation with period $T_1$ between sites $i+1$
and $i+2$. When a time interval of $T_1/2$ has elapsed
the particle has completely tunneled to site $i+2$.
The driving field can then be switched back to its original values
and the procedure repeated. This has the effect of stepping
the particle through the lattice in a sequence of
discrete moves.

\begin{center}
\begin{figure}
\includegraphics[width=0.4\textwidth,clip=true]{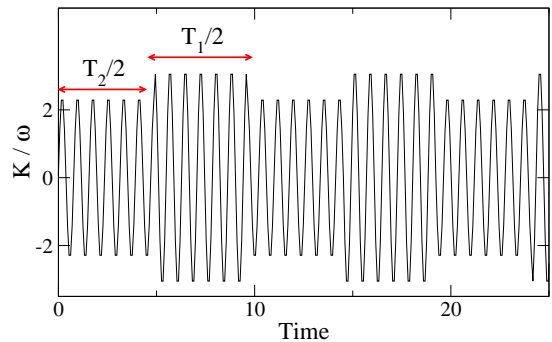}
\caption{The driving field producing
the ratchet-like motion seen in Fig.\ref{simulation}
consists of a high-frequency
sinusoidal oscillation with a modulated amplitude.
The two values of the modulation satisfy $K/\omega=2.4048/x_i$,
and act to suppress the tunneling between sites
separated by distances of $x_i$ respectively. The time-intervals $T_1$
and $T_2$ are the Rabi periods for the renormalized tunneling processes
between sites separated by $x_1$ and $x_2$.}
\label{field}
\end{figure}
\end{center}

An example of a driving field that can produce this
effect is shown schematically in Fig.\ref{field}. 
It can be thought of as a high-frequency ``carrier wave''
whose amplitude is modulated by a
squarewave envelope. The lower amplitude segment 
suppresses tunneling between sites separated by $x_1$ and has a
duration such that the particle tunnels exactly
to its other neighbor (separated by $x_2$): the reverse is true
for the higher amplitude segments. Fig.\ref{simulation}b shows
the response of the single-particle system to this
field. Instead of the two-site Rabi oscillation seen previously,
the particle now advances to the left in a series of well-defined steps.
Conversely, if the order of the modulation is interchanged,
the particle will propagate solely to the {\em right}
as shown in Fig.\ref{simulation}c. 
It is interesting to note that the direction 
of the particle's propagation also depends on which site of
the double-well it is initialized; particles started in the left well
will move in the opposite direction to those placed in the right. 
The direction of motion thus depends on both the parity
of the lattice site and the order of modulation,
in a way not seen in standard dissipative ratchets.
This flexibility requires, however, excellent control over
the localization of the initial state.

We have so far considered the extreme cases in which propagation
only occurs in one direction. If, however, the 
duration of the initial modulation is not exactly
half a Rabi-period, the initial tunneling process will not be
complete. Consequently the particle will divide into two parts, 
and under the subsequent influence of the driving field
one part will propagate to the left while the other moves to the right.
In Fig.\ref{simulation}d we show that if the
initial modulation has a duration of $T_2/4$
the particle splits in half. In this way, the
driving field can not only be used to control the motion
of a particle, but also as a quantum beam splitter to divide
a particle into a given superposition of left and right propagating
components.

We have so far just considered single-particle manipulation. However,
the ability to control the inter-site tunneling also enables us
to entangle particles and thereby realize quantum gates \cite{teich}. 
To demonstrate this, we first consider a two-site model
occupied by two particles ($a$ and $b$) that are {\em distinguishable}.
This can be realized, for example, by using
bosonic atoms \cite{jaksch_qc, jaksch}
with two different internal states \cite{internal}.
For simplicity, we model the interaction between the particles
as a Hubbard interaction
\begin{equation}
H_I = \frac{U}{2} \sum_i n_i (n_i - 1) \ ,
\label{hubbard}
\end{equation}
where $U$ sets the interaction-strength
and $n_i = n_i^a + n_i^b$ gives the total number of bosons
occupying site $i$. 
The dynamics of this system is governed by the interplay between
the kinetic energy and the interaction, and consequently
can exhibit a rather complicated time-evolution. 
If, however, $U$ is much larger than the tunneling
amplitude, the ground state of the system will then approximately
consist of each site
holding one particle, from which the doubly-occupied states
will be separated by an energy gap of $\sim U$. In this case
canonical perturbation theory can be applied
to eliminate the higher energy states, with the result
that the Hubbard interaction maps to an effective
{\em Heisenberg} term, with exchange constant given by
$\Jheis \simeq 4 \Jeff^2/U$. This mapping considerably
simplifies analysis of the system's dynamics, and reveals
that if the system is initialized in the
state $| a, b \rangle$, the two particles {\em swap} positions
after a time-interval $t = \pi/\Jheis$,
while after $t_S = \pi/2 \Jheis$
the {\em maximally entangled} state, 
$(| a, b \rangle + i | b, a \rangle)/\sqrt 2$,
is produced. Applying the Heisenberg interaction for a duration of $t_S$ 
thus realizes the \swap operation.

\begin{center}
\begin{figure}
\includegraphics[width=.4\textwidth,clip=true]{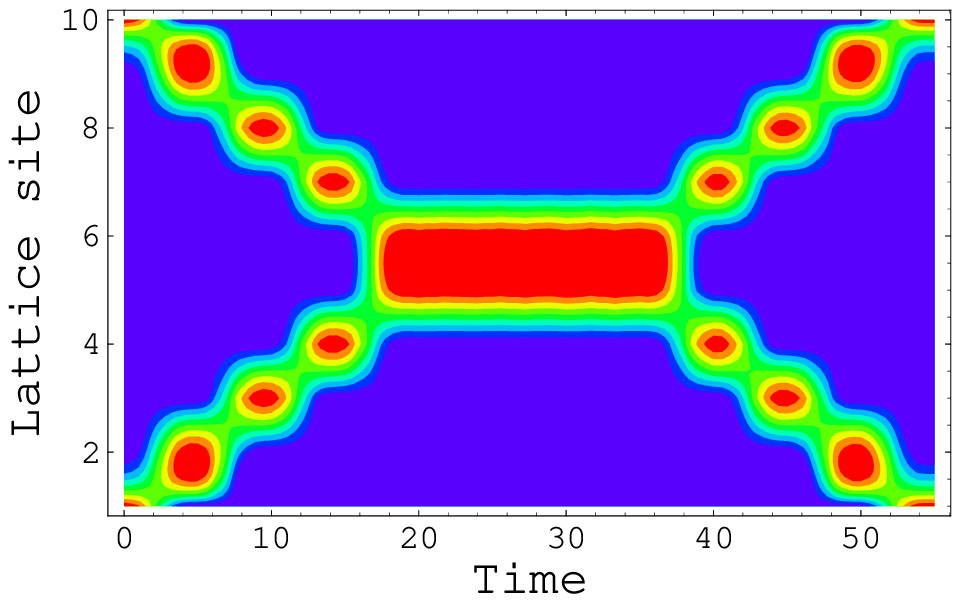}
\includegraphics[width=.4\textwidth,clip=true]{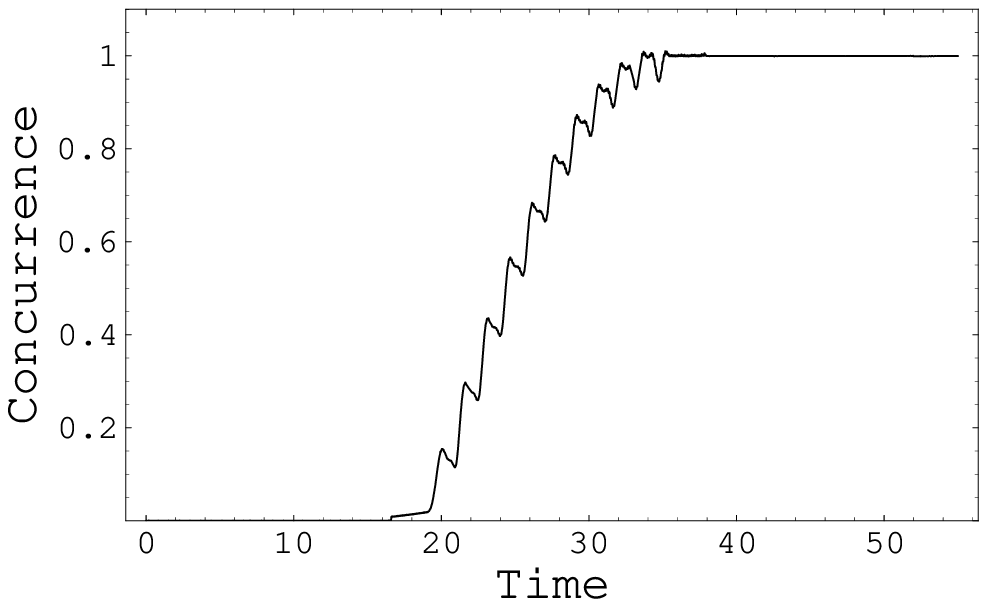}
\caption{Above: Time development of the system initialized
with one boson in the first lattice site, and another
in the final site. Under the influence of the driving field
they are moved toward the center of the array, held there
for a time-interval $t_S$, and then restored to
their original positions.
Below: The pairwise concurrence present in the system.
Initially this is zero since the initial state is not entangled,
but rises as the two bosons interact (via Heisenberg exchange).
The interval over which the interaction occurs is
chosen to maximize the concurrence, by producing
a maximally entangled state.}
\label{pair}
\end{figure}
\end{center}

In Fig.\ref{pair} we show the time-evolution of
a 10-site lattice, initialized with a boson of type $a$ in the first site,
and a boson of type $b$ in the last.
To ensure the validity of the mapping to the Heisenberg interaction,
we require a high value for $U$. However, the time required
to entangle the particles is proportional to $U$, and so to 
complete as many quantum gate operations as possible
within the system's decoherence time we would like to take
$U$ to be as small as possible. We thus consider an intermediate
value of $U=4 J$, and as before use a
driving frequency of $\omega = 32 J$ 
to place the system in the high-frequency regime.
Under the influence of the driving field 
the two bosons are progressively stepped through the lattice towards
the two central sites, whereupon the amplitude of the driving is
held at a constant value to retain the two particles
there. After being held there for a time interval of
$t_S$, the particles are then separated and returned
to the first and last lattice sites.

From Fig.\ref{pair}a it can be clearly seen that the
particles remain highly localized in space,
and accordingly the probability distribution is
generally peaked at two sites. This permits a simple measurement of
the entanglement present in the system 
by projecting the wavefunction onto just these two sites,
and then evaluating the two-particle concurrence $C(t)$ \cite{wooters},
plotted in Fig.\ref{pair}b. Initially the concurrence
is zero since the particles have not
interacted, and so the two-particle wavefunction is factorisable.
This remains true as the particles approach each other, until
they reach the two central sites. Driven by the Heisenberg interaction,
the concurrence then rises from zero following the approximate time-dependence
$C(t) = \left| \sin \Jheis t  \right|$. The high-frequency
ripples visible in this quantity arise from the influence
of the higher energy states: if $U$ is increased these ripples
will be quenched, but equally $\Jheis$ will be reduced,
and so the time-scale for entanglement to occur will increase.
When the particles are separated the degree of entanglement
remains ``frozen-in'' at its final value.
It is thus possible to generate any desired degree of entanglement
by controlling the period during which the particles interact.
When this period is equal to $t_S$, as shown in Fig.\ref{pair},
the entanglement is maximized, and the
final state of the system thus represents
a mesoscopically separated, maximally-entangled two-particle state.

\smallskip
 
{\em Conclusions}
In summary, we have shown how a periodic driving field 
can induce a novel ratchet-like motion, which
can be employed to selectively guide and divide
particles. In addition the interaction between the particles 
can be used to entangle them, and thus realize
fundamental two-qubit quantum gates, such as \swap.
In this work we have just considered a one-dimensional geometry, 
but an attractive aspect of optical lattices is the possibility
of generating a higher-dimensional \cite{twodim} lattice potentials, 
which would allow parallel processing of qubits, and thus be used 
to greatly enhance their error tolerance.
Finally, although we have specifically
considered a system of ultracold bosons, 
the method we have described could equally be applied
to optically-confined fermionic atoms \cite{fermion}, or to electronic
transport in systems such as coupled quantum dots \cite{creff_array}
or molecular wires \cite{sigmund}.

\bigskip

This research was supported by the EPSRC. The author acknowledges
the hospitality of the University of Edinburgh where this
work was completed, and thanks Sougato Bose for stimulating conversations.


\begin{thebibliography}{99}
\bibitem{internal}
{It is important to note, however, that the two species of atoms should
see the same optical lattice potential.}
\end{thebibliography}
\end{document}